\documentclass[fleqn,10pt]{wlscirep}
\usepackage[utf8]{inputenc}
\usepackage[T1]{fontenc}
\usepackage{graphicx}
\usepackage{url}

\usepackage{dblfloatfix}
\usepackage{amsmath,algorithm}
\usepackage{amsthm}
\usepackage[noend]{algpseudocode}
\usepackage{setspace}
\usepackage{color}
\usepackage{tikz}
\usepackage{pdfpages}
\usetikzlibrary{arrows}
\usepackage{algorithm}
\usepackage{cases}
\usepackage{relsize}
\usepackage{tablefootnote}
\usepackage{threeparttable}
\usepackage{booktabs}
\usepackage{makecell}
\usepackage[markup=hmarkupi]{changes}
\usepackage{amsmath}
\usepackage[noend]{algpseudocode}
\usepackage{cancel}
\usepackage{algpseudocode}
\algnewcommand{\Inputs}[1]{%
  \State \textbf{Inputs:}
  \Statex \hspace*{\algorithmicindent}\parbox[t]{.8\linewidth}{\raggedright #1}
}
\algnewcommand{\Initialize}[1]{%
  \State \textbf{Initialize:}
  \Statex \hspace*{\algorithmicindent}\parbox[t]{.8\linewidth}{\raggedright #1}
}

\usepackage{algorithm}
\usepackage{algpseudocode}

\usepackage{amssymb}

\makeatletter
\def\BState{\State\hskip-\ALG@thistlm}
\makeatother

\usepackage[T1]{fontenc}
\usepackage[font=small,labelfont=bf,tableposition=top]{caption}

\DeclareCaptionLabelFormat{andtable}{#1~#2  \&  \tablename~\thetable}

\usepackage{amsmath}

\newcommand{\RN}[1]{%
  \textup{\lowercase\expandafter{\romannumeral#1}}%
}

\usepackage{graphics}

\usepackage{blindtext}
\usepackage{enumitem}
\usepackage{xcolor}
\usepackage{graphicx}
\usepackage{caption}
\usepackage{ragged2e}
\usepackage[font=small,labelfont=bf,
   justification=justified,
   format=plain]{caption}

\usepackage[T1]{fontenc}
\usepackage[utf8]{inputenc}
\usepackage{babel}
\usepackage[font=small,labelfont=bf]{caption}

\usepackage{tabularx}

\usepackage{lipsum}
\usepackage{mwe}
\usepackage{geometry}
\usepackage{listings}
\usepackage{caption}
\usepackage[export]{adjustbox}
\usepackage{tcolorbox}
\usepackage{multirow}
\usepackage{subcaption}
\usepackage{ragged2e}
\usepackage[T1]{fontenc}
\usepackage[utf8]{inputenc}
\usepackage{babel}
\usepackage[font=small,labelfont=bf]{caption}
\usepackage{booktabs}
\usepackage{amsmath}
\usepackage{multicol}

\usepackage[T1]{fontenc}
\usepackage[english]{babel}
\usepackage{algorithm,algorithmicx,algpseudocode}
\usepackage{amssymb}

\usepackage{calligra}
\DeclareFontFamily{T1}{calligra}{}
\DeclareFontShape{T1}{calligra}{m}{n}{<->s*[1.1] callig15}{}
\DeclareMathAlphabet{\mathcalligra}{T1}{calligra}{m}{n}

\usepackage{mathrsfs}

\algnewcommand{\algorithmicand}{\textbf{ and }}
\algnewcommand{\algorithmicor}{\textbf{ or }}
\algnewcommand{\OR}{\algorithmicor}
\algnewcommand{\AND}{\algorithmicand}
\algnewcommand{\var}{\texttt}

\newcommand{\M}[1]{{\bf #1}}
\newcommand{\V}[1]{{\bf #1}}
\DeclareFontFamily{OT1}{pzc}{}
\DeclareFontShape{OT1}{pzc}{m}{it}{<-> s * [1] pzcmi7t}{}
\DeclareMathAlphabet{\set}{OT1}{pzc}{m}{it}

\newcommand{\T}[1]{\textrm{#1}}

\title{The impact of input node placement in the controllability of brain networks}

\author[1]{Seyed Samie Alizadeh Darbandi}
\author[2]{Alex Fornito}
\author[1,*]{Abdorasoul Ghasemi}
\affil[1]{Dept. of Computer Engineering, K. N. Toosi University of Technology, Tehran, Iran}
\affil[2]{The Turner Institute for Brain and Mental Health, School of Psychological Sciences, and Monash Biomedical Imaging, Monash University, Melbourne, VIC, Australia}

\affil[*]{corresponding.author arghasemi@kntu.ac.ir}

\begin{abstract}
Network control theory can be used to model how one should steer the brain between different states by driving a specific region with an input. The needed energy to control a network is often used to quantify its controllability, and controlling brain networks requires diverse energy depending on the selected input region. We use the theory of how input node placement affects the longest control chain (LCC) in the controllability of brain networks to study the role of the architecture of white matter fibers in the required control energy. We show that the energy needed to control human brain networks is related to the LCC, i.e., the longest distance between the input region and other regions in the network. We indicate that regions that control brain networks with lower energy have small LCCs. These regions align with areas that can steer the brain around the state space smoothly. By contrast, regions that need higher energy to move the brain toward different target states have larger LCCs. We also investigate the role of the number of paths between regions in the control energy. Our results show that the more paths between regions, the lower cost needed to control brain networks. We evaluate the number of paths by counting specific motifs in brain networks since determining all paths in graphs is a difficult problem.
\\
\textbf{Keywords: } complex systems, brain networks, structural controllability, control energy
\end{abstract}
\begin{document}

\flushbottom
\maketitle
%
%
\thispagestyle{empty}

\section{Introduction}
The brain is a complex system constructed from billions of neurons linked by complicated patterns of white matter fibers \cite{deco2011emerging,hagmann2008mapping,mivsic2015cooperative}. Cognitive function results from dynamical transitions of the neurophysiological activity in support of complex behaviours\cite{cocchi2013dynamic}. Recognizing the mechanisms and processes of these transitions by controlling the dynamics of brain networks is critical for many applications like discovering how the system can be altered by disease states, intervening in disease to redirect changes\cite{deco2014great}, development\cite{cao2014topological}, and rehabilitation\cite{weiss2011functional}.

We can consider a spatially contiguous ensemble of neurons as a region. While the dynamics of cognitive functions throughout these regions are almost nonlinear, the processes can be approximated by linearized generalizations\cite{galan2008network} that are useful to characterize how the anatomical structure of the brain influences its dynamic functions\cite{betzel2016optimally,becker2018spectral}. In this respect, a brain can be represented as a graph in which regions form the graph nodes, and anatomical connections create the links between them\cite{bullmore2009complex, bassett2018nature}. See Ref. \cite{fornito2016fundamentals} for more details of modeling brain networks as graphs. According to the analysis\cite{ruiz2014real, fox2012measuring}, we can change the activity state of neurons in a region by external signals such as electrical stimulation\cite{muldoon2016stimulation} or task modulation\cite{cui2020optimization}. This change can then alter the state of other brain regions. The concept of controllability addresses how we can steer the state of a network from its initial state to a desired one by changing the state of a sufficient number of network nodes by appropriate external signals.

The first question in investigating brain controllability concerns how many and which nodes should receive external signals to control the network. Regions whose state in controllability is changed by external signals are called input nodes, and the smallest required number of external signals to ensure controllability can be easily determined by solving a graph-based problem called maximum matching \cite{liu2011controllability,commault2013input,ruths2014control}. For a more technical introduction and the relationship between controllability and structural features of networks, e.g., the degree distribution \cite{ghasemi2020diversity}, see Refs.\cite{tang2018colloquium,bassett2017network, wu2019controllability,menara2018structural}. Brain networks need only a few input nodes to guarantee controllability\cite{gu2015controllability, menara2018structural}. In particular, brain networks are theoretically controllable by one region as an input node, and we can move the brain state from the initial state to a desired one by applying a proper signal to any region as an input region.

However, determining the required input nodes to ensure controllability is just a theoretical approach, and it does not provide any information about the energy required for controlling the network in reality. In practice, the energy required for brain controllability following input to a single region may be unrealistic\cite{yao2017functional,gu2015controllability}. This is a critical consideration, given the high metabolic expense of brain functions. The required energy to control brain networks can also vary across different input regions \cite{gu2015controllability}. Understanding the energetic cost of brain controllability is thus essential to developing plausible strategies for directing brain activity toward desired states, such as in brain therapies. Here three questions arise: (i) what are the energy requirements of different input regions to establish controllability, (ii) what is the reason for regional differences in the required energy, and (iii) how can we reduce the required control energy?

Work addressing these questions has identified two classes of nodes. Regions high in average controllability can, on average, steer the system into different target states with little control energy, whereas regions with high modal controllability correspond to areas that push the brain in states that are energetically difficult to reach\cite{gu2015controllability,deng2020controllability}.
Considering the effect of external signals and other parameters on the required energy has also been the subject of some studies\cite{yao2019toward,gu2017optimal,karrer2020practical}.
However, while these studies shed light on which regions show different control energy requirements, they do not address the reason for this difference or consider how we might reduce those energetic requirements. Furthermore, using only control theory methods to investigate these two remaining questions leads to NP-hard problems that are difficult to solve \cite{tzoumas2015minimal,alizadeh2023input}. Hence, we need to employ other measures to investigate more about the energy in brain networks.

In this study, we show how the structure of white matter fibers affects the energy required for regional controllability using the concept of the Longest Control Chain (LCC) in the network, i.e., the longest distance between input and non-input nodes.
Numerically and theoretically, it's shown that the control energy strongly correlates with the LCC in the network\cite{chen2016energy, klickstein2018control, chen2015paradox}. We also have indicated how input node placement can modify the LCC in the network to reduce the required energy of controllability, so the network structure and the location of inputs play a significant role in determining the control cost\cite{alizadeh2023input}. Here, we leverage this result to provide an insight into the relationship between the regional differences in the needed energy for controlling brain networks and the architecture of white matter fibers. Additionally, we build on our prior work to suggest a solution to decrease the needed energy to control brain networks by applying additional external signals to the minimum number of regions to reduce the LCC in the network. Our work thus provides a formal bridge that links white matter architecture, controllable dynamics, and the energetics of brain function to study the difference in regional control energy and to reduce the cost of brain networks controllability.

\section{Theoretical framework and data}

\subsection{Network controllability and control energy} \label{Network controllability and control energy}

Network control theory offers a promising mathematical framework for linking the structure of a network to the dynamic it can support. To formally define the model, consider a directed network $\set G(\set{V}, \set{E})$ where $\set V=\{v_i | i=1,..., N\}$ is the set of nodes and $\set E \subset \set V \times \set V$ is the set of weighted directed links, i.e., each link $(v_i\rightarrow v_j)$ has an associated weight $a_{ji}\in \mathbb R$ representing the strength of the interaction.
We assign a state $x_i\in \mathbb R$ to each node $v_i$ which evolves following the equation
\begin{equation}
\dot{\V x}(t) = \M A \V x(t) + \M B \V u(t), \label{eq: LTI}
\end{equation}
where the first term represents the internal dynamics of the system and the second term expresses the external control imposed on the network.
Specifically, $\V x={[x_1, x_2, ..., x_N]}^T$ is the state of the system as a vector of all node states, $\M A \in \mathbb R^{N\times N}$ is the weighted adjacency matrix that shows the interaction between network nodes, $\V u(t)={[u_1, u_2, ..., u_M]}^T\in \mathbb R^M$ is a vector of $M$ time-dependent control signals, and matrix $\M B\in \mathbb R^{N\times M}$ defines how the control signals are coupled to the system. In controllability of human brains, the state of the system $\V x(t)$ reflects the magnitude of the neurophysiological activity of the $N$ brain regions at a specific time, control signals $\V u(t)$ correspond to different inputs that change the states of the system such as electrical stimulation or task modulation, the input matrix $\M B$ indicates input regions whose states are modified by external signals, and adjacency matrix $\M A$ encodes the white matter connections between different brain regions as derived using methods detailed in the next section (Sec. (\ref{brain_network_construction})).

\begin{figure}[!h]
\begin{subfigure}[h]{.33\textwidth}
\centering
\includegraphics[width=.93\linewidth]{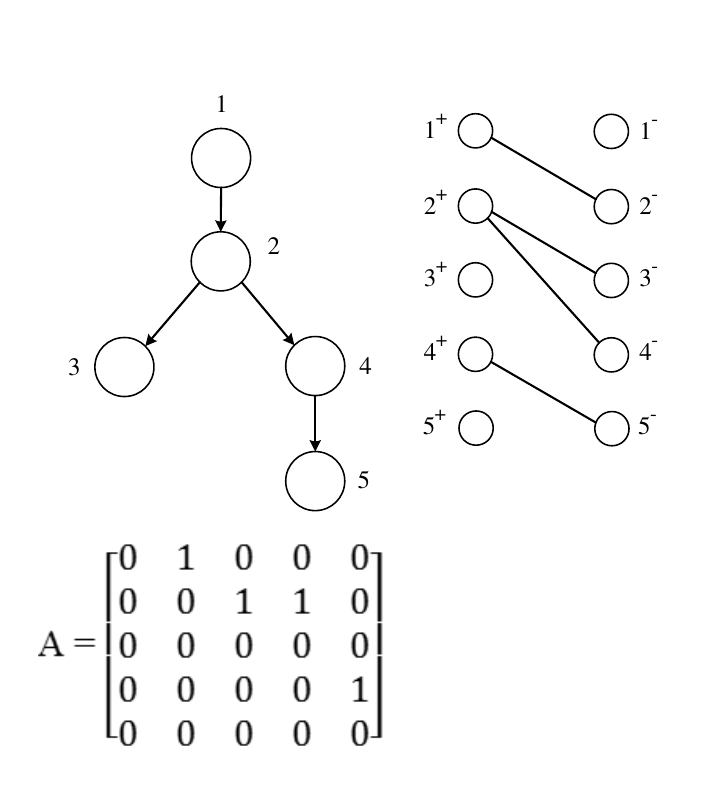}
\caption{}
\label{fig: Fig-1 (a)}
\end{subfigure}
\begin{subfigure}[h]{.33\textwidth}
\centering
\includegraphics[width=.93\linewidth]{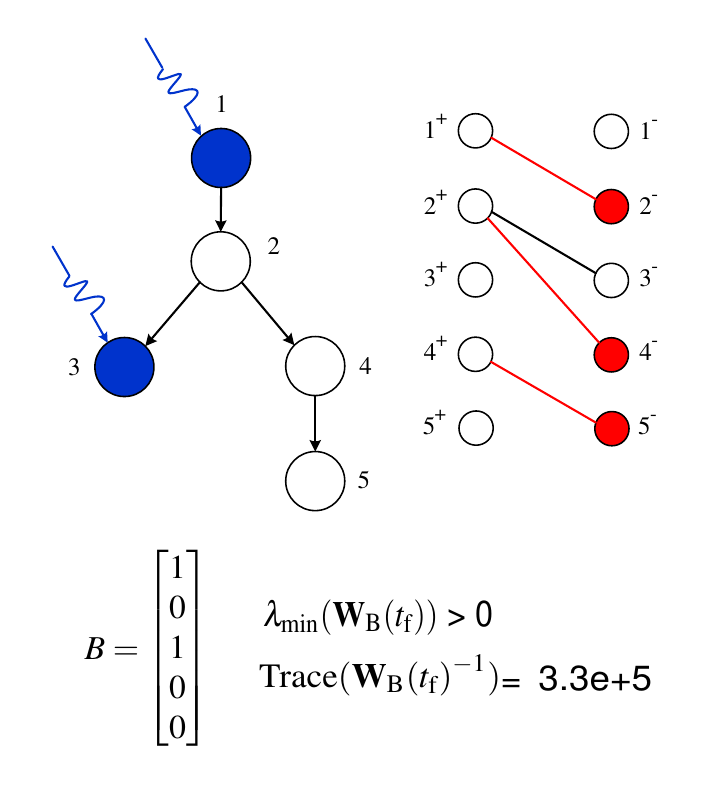}
\caption{}
\label{fig: Fig-1 (b)}
\end{subfigure}
\begin{subfigure}[h]{.33\textwidth}
\centering
\includegraphics[width=.93\linewidth]{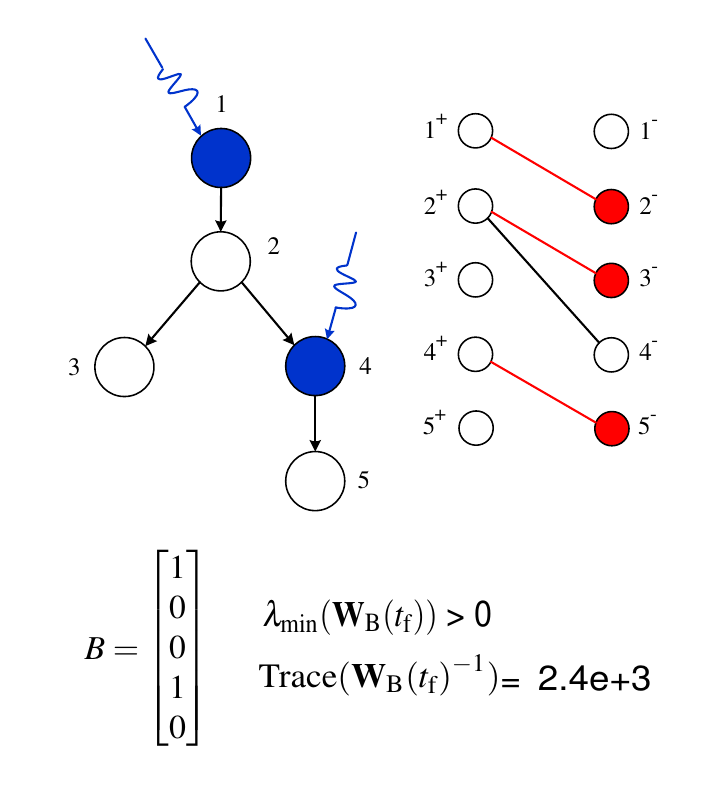}
\caption{}
\label{fig: Fig-1 (c)}
\end{subfigure}
\caption{\textbf{The structural controllability and the effect of LCC on the control energy. } \textbf{(a)} the directed network $\set G$, its bipartite graph $\set B$, and the structural adjacency matrix $\M A$ of $\set G$. \textbf{(b)} a controllability scheme in which the matching set $\set E_\T{M} = \{(1^+,2^-),(2^+,4^-),(4^+,5^-)\}$ is shown by red links in the corresponding bipartite graph. Also, matched and unmatched nodes in the set $\set V^-$ are shown in red and white, respectively. The $\set E_\T{M}$ is the maximum number of links for which no two links share the same start or end node. Here, nodes 1 and 3 are unmatched nodes to apply independent external signals to control the network. In this scheme, the distance of network nodes from the closest input is [0, 1, 0, 2, 3], respectively, so the LCC equals three. \textbf{(c)} As the maximum matching implementation is not unique, we selected another possible matching set in which $\set E_\T{M} = \{ (1^+,2^-),(2^+,3^-),(4^+,5^-)\}$. Here, the number of unmatched nodes is the same $N_\T{u}=2$, but inputs are nodes 1 and 4. Also, the distance between inputs and network nodes is [0, 1, 2, 0, 1], respectively, so the LCC equals two. As is shown, $\text{Trace}({\M W_\textrm{B}(t_\T f)}^{-1})$ in \textbf{(c)} is smaller than \textbf{(b)} by two orders of magnitude showing the control scheme \textbf{(c)}, which has smaller LCC, needs less control energy than other one with the same number of input nodes but larger LCC.}
\label{fig: Fig-1}
\end{figure}

A system $(\M A,\M B)$ is controllable if, with the proper choice of control signals, we can drive it from any initial state $\V x(0)$ to any final state $\V x(t_\T f)$ in finite time. Determining controllability, in general, is a numerically unstable problem that requires exact knowledge of the link weights, making it difficult to study real networks directly. In brain networks, the information about the link weights is either unknown or inferred with noisy estimates, which is a barrier to studying their controllability. Structural controllability focuses on the structure, not the weights, and it implies that we can investigate the controllability of brain networks by only knowing the links between regions without their strengths\cite{lin1974structural}. So, in the adjacency matrix $\M A$, if an element $\M a_{ij} \neq 0$, we substitute it to one to study structural controllability. It is shown that if a network is structurally controllable, it is controllable for most weight assignments to links\cite{lin1974structural,liu2011controllability}. Results in control theory show that network $\M A$ given a set of input nodes is controllable if the smallest eigenvalue of the controllability Gramian matrix $\lambda_\T{min}(\M W_\textrm{B}(t_\T f))$ is larger than zero where ${\M W_\textrm{B}}(t_\T f)=\int_{0}^{t_\T f} e^ {\M A \tau} \M B \M B^T e^ {\M A^T\tau}d\tau$ ~ \cite{rughlinear,yan2015spectrum,gu2014controllability}. Finding the minimum required input nodes to ensure structural controllability is an NP-complete problem\cite{olshevsky2014minimal}, however, if a control signal can couple to multiple nodes, the minimum number of external signals to ensure structural controllability can be determined in polynomial time by the maximum matching algorithm in the bipartite graph of $\set G$. The bipartite representation $\set B$ of a directed network $\set G$ can be shown by splitting each node $v\in \set V$ into two copies $v^+\in \set V^+$ and $v^-\in\set V^-$. If there exists a directed link $(v\rightarrow w)$ in $\set G$, we add an undirected link $(v^+ - w^-)$ to $\set B$. A matching in the bipartite graph $\set B (\set V^+ \cup \set V^-, \set E')$ is a subset of links $\set E_\T M \subset \set E'$ such that no two links in $\set E_\T M$ share start or endpoints, and a node in the set $\set V^-$ is unmatched if no link in $\set E_\T M$ is pointing at it. See figure (\ref{fig: Fig-1}) for an illustrative example. Therefore, if $N_\T i$ shows the number of external signals and $N_\textrm{u}$ indicates the number of unmatched nodes in the maximum matching, then
\begin{equation}
N_\textrm{i} = max (1,N_\textrm{u}). \label{eq: N_i}
\end{equation}

\begin{figure}[!h]
\centering
\includegraphics[width=.4\linewidth]{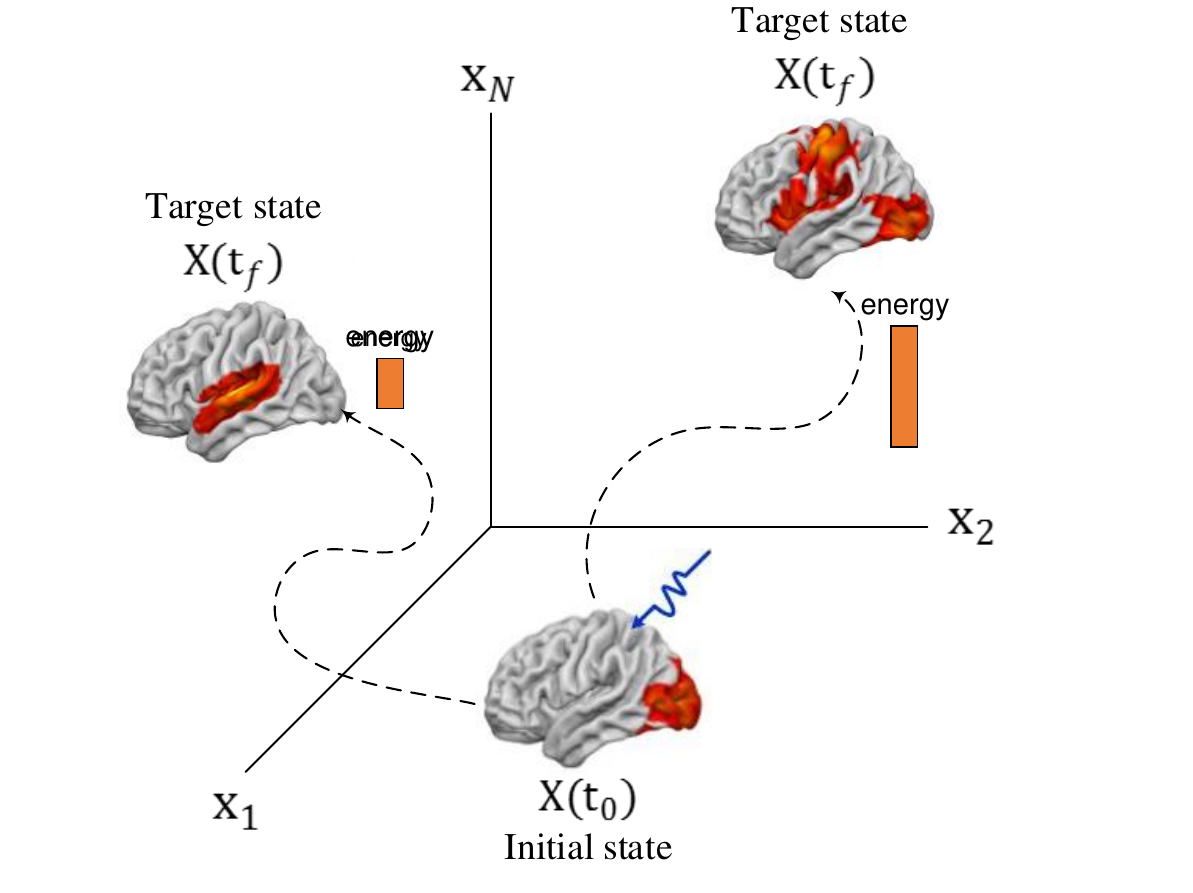}
\caption{\textbf{The controllability of brain networks. } As is shown, a brain's state can be moved from its initial state $X(t_\T{0})$ to different final states $X(t_\T{f})$ by applying a proper external signal to a given region. The required control energy in steering the brain state to different target states could be varied, and we evaluate the average required energy over all possible target states to analyze the control energy of an input region. }
\label{fig: Fig-2}
\end{figure}

Solving the maximum matching problem provides a set of nodes to control the system; however, it does not provide any information about the required energy to control the system. The energy of controllability given a set of input nodes can be evaluated by different metrics. For example, one may consider the average required energy to steer the network to all possible final states, which can be calculated using the controllability Gramian matrix, ${E(t_\T f)} = \text{Trace}({\M W_\textrm{B}(t_\T f)}^{-1})$\cite{muller1972analysis}. We use $\text{Trace}({\M W_\textrm{B}(t_\T f)})$ instead of $\text{Trace}({\M W_\textrm{B}(t_\T f)}^{-1})$ as a measure of control energy for two main reasons: first, $\text{Trace}(\M W_\textrm{B}(t_\T f)^{-1})$ and $\text{Trace}(\M W_\textrm{B}(t_\T f))$ satisfy a relation of inverse proportionality\cite{summers2014optimal}, $\text{Trace}(\M W_\textrm{B}(t_\T f)^{-1}) > N^2/\text{Trace}(\M W_\textrm{B}(t_\T f))$, so that the information obtained from the two metrices are correlated with one another and, second, ${\M W_\textrm{B}}(t_\T f)$ is typically ill-conditioned and close to singularity, so $\text{Trace}(\M W_\textrm{B}(t_\T f)^{-1})$ cannot be accurately computed even for small brain networks. Therefore, we use
\begin{equation}
\varepsilon(t_\T f) = \text{Trace}({\M W_\textrm{B}(t_\T f)}), \label{eq: average_energy}
\end{equation}
to evaluate the control energy of brain networks, with higher values indicating that less energy is required to move around the state space in all directions (see figure (\ref{fig: Fig-2})). However, employing and calculating these parameters to regulate the control energy not only leads to an NP-complete problem but does not provide any extra information about the reason for having high or less control energy and can be used only to evaluate the control effort numerically.

It has been shown that control energy can be evaluated by the structural parameter longest control chain (LCC), i.e., the longest distance between non-input nodes and the closest input to them
\begin{equation}
\T{LCC} = \max_{w\in\set V}\min_{v \in \set S} d(v,w), \label{eq: LCC}
\end{equation}
where $\set S$ is the set of input nodes and $d(v,w)$ is the length of the shortest path connecting from $v$ to $w$. It is proven that any change in LCC can significantly reduce or increase the energy necessary to control networks\cite{chen2016energy,alizadeh2023input, LCC,klickstein2018control,chen2020optimizing}. Figure (\ref{fig: Fig-1}) shows an illustrative example of how LCC affects the control energy in a small network. See \cite{alizadeh2023input} for more details. 

In this work, we study brain networks in which regions construct network nodes, and the white matter fibers create links connecting nodes. Then, we investigate the controllability of brain networks under the dynamics specified in equation (\ref{eq: LTI}). We need three parameters, adjacency matrix $\M A$, input matrix $\M B$, and the control time $t_\T{f}$. The way of constructing the adjacency matrix $\M A$ of brain networks will be described in the next section, and the control time is considered $t_\T{f} = 1$. The input matrix $\M B$ can be constructed by setting the node index of the input region to one. Then, employing the input node placement and its effect on the LCC, we investigate the controllability of brain networks using $\lambda_\T{min}(\M W_\textrm{B}(t_\T f))$ to examine the controllability, $\varepsilon(t_\T f)$ to evaluate the control cost, and equation (\ref{eq: LCC}) to determine the LCC.

\subsection{Data acquisition and brain network construction} \label{brain_network_construction}

Diffusion tensor imaging (DTI) is an MRI technique that uses anisotropic diffusion to estimate the white matter organization of the brain. In this method, sets of neurons are assigned to regions whose total number could be in diverse scales. Here, we use Automated Anatomical Labeling (AAL)\cite{tzourio2002automated}, an atlas of the human brain in which neurons are assembled into 90 regions. Therefore, network nodes in constructed brain networks in this study present regions of the AAL atlas. This architecture considers the anatomical and functional characteristics of brains\cite{tzourio2002automated}. Consistent with prior works\cite{yao2019toward}, it leads to a matrix $\in$ $R^{90 \times 90}$, which displays the adjacency matrix of brain networks. The obtained result by processing the DTI outcome is a weighted adjacency matrix. In the first step, we set its diagonal entries to zero to eliminate the self-loops in the network. Then, we employ a threshold of the connection weights, causing some links to be removed, i.e., setting some of its non-diagonal entries to zero. In this respect, a link between two regions only exists if its connection weight is more than the defined threshold. Eventually, we summarize these extracted estimations of links between 90 regions in a weighted adjacency matrix that shows the connection between different regions in studied networks while regions would be mapped to the AAL atlas. More details about the data collecting setup and parameters for DTI are provided in the section \ref{secs: methods}. We can modify the weighted adjacency matrix to a binary matrix by setting each element $a_{i j} > 0$ to one to study the structural controllability. The corresponding structural adjacency matrix focuses on the brain network connections, not their weights. Furthermore, this adjacency matrix represents the brain with a symmetric structure which gives information about the links, not their directions. However, any unidirectional link between two nodes $i$ and $j$ could be considered a bidirectional link in such a way that there is a connection from $i$ to $j$ and oppositely from $j$ to $i$. Since we employ the control theory framework on directed networks, we considered this interpretation to construct the directed brain networks.

\section{Methods}\label{secs: methods}

In this work, we considered nine healthy human adults in different ages and genders to investigate their brain networks. As mentioned in the section (\ref{brain_network_construction}), we first use the MRI technique to scan their brains. Data collecting setups and parameters are as the following. 

\subsection*{Image acquisition}
MRI images were acquired on a Siemens 3.0 Tesla scanner (Prisma, Erlangen, Germany), at the Iranian National Brain Mapping Lab (NBML) (www.nbml.ir). A few characteristics of this machine included 50-cm FOV with the industry best homogeneity; whole-body; superconductive zero helium oil-off 3T magnet; and head/neck 20 direct connect.  We used the 64-channel head coil in our study. The MRI protocols were selected to match the international projects, such as the UK Biobank or the ENIGMA consortium.
Using an EPI Diffusion weighted protocol, 66 volumes were acquired with the imaging parameters of TR= 9900 (ms), TE= 90 (ms), b-value= 0 (s/mm²), b-value= 1000 (s/mm²), diffusion directions = 64; TA= 11:05 min, voxel size = 2.0×2.0×2.0 (mm), FOV read =256 (mm); number of slices= 65; distance factor = 0; phase encoding direction= anterior >> posterior, matrix size = 128×128×65.
Using an MPRAGE protocol, T1-weighted structural scans were acquired for clinical diagnostic purposes, with the imaging parameters of TA = 4:12 min, TR = 1800 (ms), TE = 3.53 (ms), TI = 1100 ms, flip angle = 7 degrees, voxel size = 1.0×1.0×1.0 (mm), multi-slice mode = sequential, FOV read = 256 (mm), number of slices = 160, phase encoding direction = anterior >> posterior, matrix size = 256×256×160.

\subsection*{DWI data Analysis Method}
DWI data analysis and tractography were performed with ExploreDTI software 
, University Medical Center, Utrecht, the Netherlands in the following order: 
\begin{itemize}
\item {1.	DWI data pre-processing}: (i) data conversion from DICOM formats to NIfTI and Bmatrix text-file formats; (ii) the right-left orientation was checked using the flip/ permute plugin; (iii) conversion to MAT files; (vi) the orientation of the MAT files was checked using the glyph plugin; (v) correction for subject motion an and eddy current/EPI distortions using a cubic spline interpolation.
\item{2.	Tractography}: We use a deterministic streamline method to obtain fibers, by setting the Fractional Anisotropy (FA) threshold to 0.2, the angular threshold to 30 (degree), linear interpolation method and fiber length range:50-500 (mm).
\item{3.	Connectivity matrix}: For network analysis we use an atlas template/labels method that obtain connectivity matrics (CMs) from AAL90 atlas \cite{tzourio2002automated}; for more information and example studies about the network analysis tools of ExploreDTI see the work of Reijmer YD et al \cite{reijmer2013disruption, reijmer2013utrecht, reijmer2015structural}. We use CMs for number of tracks in graph analysis.
\end{itemize}

\begin{figure}[!h]
\centering
\includegraphics[width=1\linewidth]{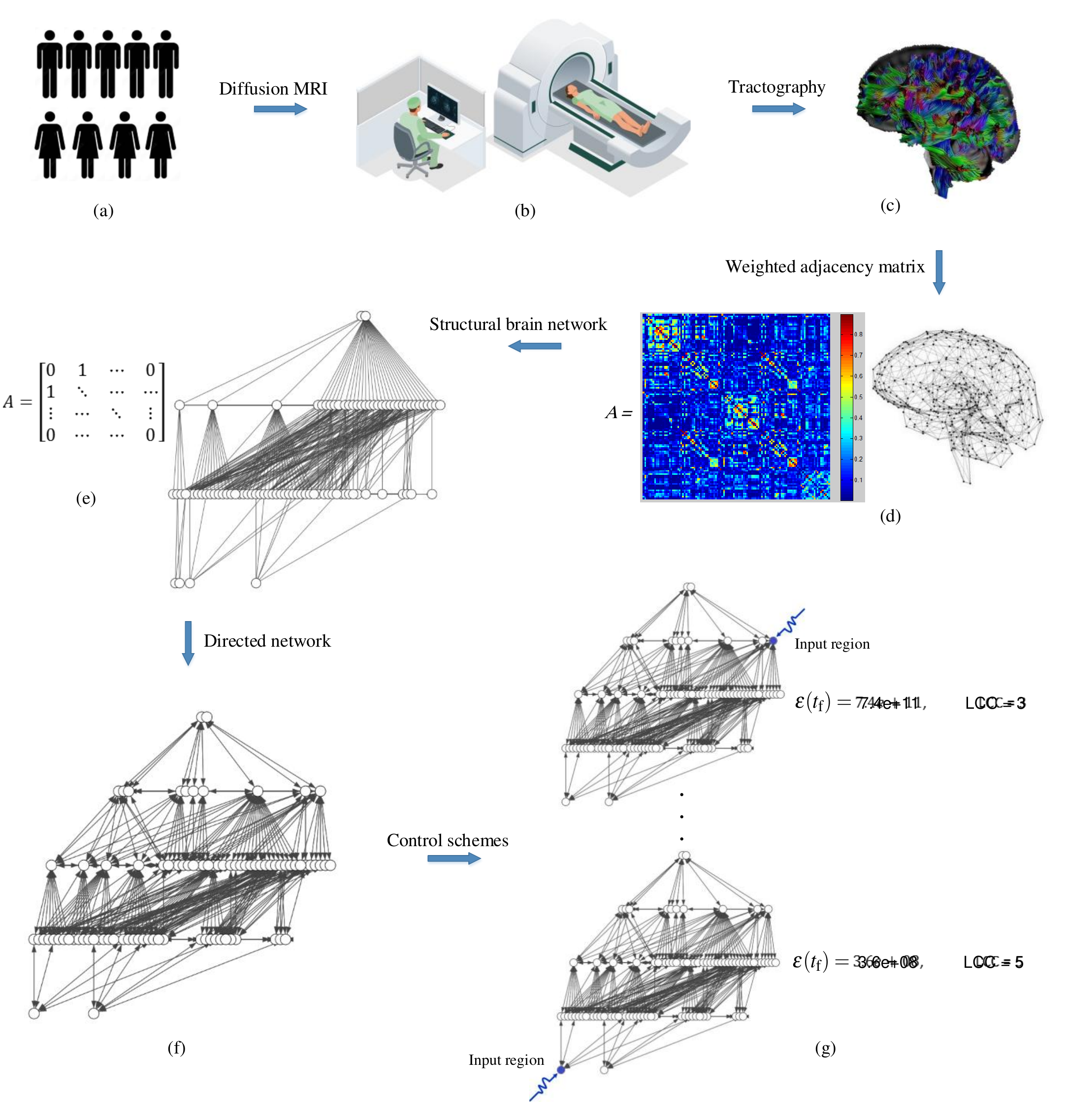}
\caption{\textbf{The schematic of the work process. }
\textbf{(a)} We used nine healthy people of different genders and ages to investigate the controllability of brain networks. \textbf{(b)}-\textbf{(c)} We used the Diffusion Tensor Imaging (DTI) scan, which offers a non-invasive look into how white matter, the highways of our brain, connect distant brain regions determined by the AAL atlas. Each fiber represents many thousands of actual fibers in the human brain. In fiber tractography imaging, the convention for color coding is as follows: red: transverse fibers, green: anteroposterior fibers, and blue: craniocaudal fibers. \textbf{(d)} the weighted adjacency matrix obtained by processing data from DTI. \textbf{(e)} We used a threshold to remove links whose weight ($a_{ij}$) is less than the threshold. Then, we set any element $a_{ij}>0$ to one to obtain the structural brain network. The illustrated network is associated with one of the real data used in this work. \textbf{(f)} we substitute any undirected link $(i, j)$ in the network to two directed links $(i \to j)$ and $(j \to i)$. \textbf{(g)} Finally, we analyze different control schemes in which any region is considered an input region. Recall that higher value for $\varepsilon(t_\T f)$ shows less required energy to move around the state space in all directions.}
\label{fig: Fig-3}
\end{figure}

After constructing the structural network of nine studied brains, we investigate their controllability concerning the created LCC. Specifically, in each brain network, we consider any region as an input node to examine different control schemes. Then, we evaluate the role of the location of input regions and the longest control chain of the controllability in the required control energy. Furthermore, we will investigate the effect of the pattern of white matter fibers between regions on the control cost. See figure (\ref{fig: Fig-3}) for a schematic of our work process.

We have shown the network properties of nine studied brain networks in the table (\ref{table: Brains_data}) in which $|\set V|$ and $|\set E|$ indicate the number of network nodes and links, respectively. Note that using the threshold of the connection weights in constructing the adjacency matrix may lead to having some isolated nodes consistent with previous works\cite{yao2017functional}. We determined these regions in each network, shown in the column $R_\T{isolated}$ of the table (\ref{table: Brains_data}). There are structural network properties that affect both network controllability and control energy. For example, we have shown previously that network density and the degree of heterogeneity are two network properties with significant impact on LCC\cite{alizadeh2023input}. In principle, the heterogeneity of a network measures the diversity in the node degrees with respect to a completely homogeneous network of the same number of nodes. We calculate the heterogeneity of nodes based on the maximum heterogeneity of input degree and heterogeneity of output degree $H = max(H^{in}, H^{out})$ where $H^{in/out}$ can be defined by $H_{in/out}=\frac{1}{rN^2}\sum_i \sum_j (k_i^{in/out}-k_j^{in/out})$\cite{liu2011controllability}. Here, $r$ is a constant, and $k_i^{in/out}$ shows the input/output degree of node $i$. We determined these network properties in each studied network that columns $c$, $K_\T{min}$, $K_\T{max}$, and $H$ show the average degree, the minimum and the maximum degree in the network, and the degree of heterogeneity, respectively. The result shows that brain networks are dense, and the degree of network nodes could be varied by a large difference. Note that in brain networks, the out-degree and in-degree are equal for each region, $K^\T{out} = K^\T{in}$.
Since LCC points to the distance between inputs to other network nodes, we also calculated the average distance between network nodes that column $d$ shows the result.

\begin{table}[h]
\centering
\small
\caption{The properties of studied brain networks}
\begin{tabular}{ll | cc | lllllccc}
\multicolumn{2}{c|}{Networks}   & \multicolumn{1}{c}{Age} & \multicolumn{1}{c|}{Gender} & \multicolumn{1}{c}{$|\set V|$} & \multicolumn{1}{c}{$|\set E|$} &  \multicolumn{1}{c}{$R_\T{isolated}$} & \multicolumn{1}{c}{$H$} & \multicolumn{1}{c}{$c$} & $K_\T{min}$ & $K_\T{max}$ & \multicolumn{1}{c}{$d$} \\ \hline  \hline
\multirow{9}{*}{} 
&1&50& F  & 90 & 1251 &   Amygdala R                                              & 0.61  &  13.9  & 1 & 35 & 2.1                         \\
&2&66& F  & 90 & 1183 &   Amygdala L - Heschl R - Rolandic Oper L & 0.54  &  13.1  & 1 & 25 & 2.3                    \\
&3&67& M  & 90 & 1220 &  -                                                                & 0.53  &  13.5  & 1 & 36 & 2.2 \\
&4&21& M & 90 & 1091 &  Amygdala R                                              & 0.62  &  12.1  & 2 & 33 & 2.2 \\
&5&31& M  & 90 & 1056 &  Amygdala R - Heschl L - Rectus L            & 0.70  &  11.7  & 1 & 45 & 2.1      \\
&6&64& F  & 90 & 1086 &   -                                                                & 0.57  &  12.0  & 1 & 26 & 2.3                         \\
&7&47& M  & 90 & 1111 &   Amygdala R - Amygdala L - Heschl L        & 0.70  &  12.3  & 2 & 37 & 2.2                         \\                    
&8&66& M  & 90 & 1120 &   Amygdala L                                               & 0.61  &  12.4  & 1 & 43 & 2.2 \\                      
&9&35& M  & 90 & 1090 &   Amygdala R                                              & 0.61  &  12.1  & 1 & 30 & 2.3 \\                     
\hline
\end{tabular}
\label{table: Brains_data}
\end{table}

\section{Results}

\subsection{The controllability of studied brain networks}
We collected and used constructed brain networks to study the controllability of human brains concerning the structural parameter LCC. We first investigate how many input nodes are required to ensure structural controllability of each network. For this purpose, we removed isolated nodes which do not affect network connections and focused on the connected graph. We then constructed the associated bipartite graph of each network and implemented the maximum matching algorithm to determine unmatched nodes as the minimum required control signals to ensure controllability. The result shows that all nodes are matched after the MM implementation, indicating the need for one control signal to ensure structural controllability, $N_\textrm{i} = max (1, N_\textrm{u}) = 1$. Therefore, theoretically, one control signal is required to steer the network's state to any desired one. We then considered each region as an input region and evaluated the smallest eigenvalues of the controllability Gramian matrix $\lambda_\T{min}(\M W_\textrm{B}(t_\T f))$. These values were greater than zero, indicating that the system is theoretically controllable through any input region. Therefore, brain networks can be moved into an arbitrary target state (for example, active memory retrieval versus mathematical calculations or health versus disease) by changing the activity of a single brain region. This result is consistent with previous studies showing brain networks need only an input node to be controllable\cite{gu2015controllability,menara2018structural}.

\subsection{Input region placement and the effect of LCC on the control energy} \label{LCC}

We now investigate the role of different regions in the controllability of brain networks concerning the required control energy and their LCC. For this purpose, we determine how the needed energy could be different in controlling brain networks. In the following analyses, all control energies are measured by $\varepsilon(t_\T{f}) = \text{Trace}({\M W_\textrm{B}(t_\T f)})$ such that a smaller value requires more control energy, on average, to control the network, as mentioned in Sec. (\ref{Network controllability and control energy}). We considered each region as an input node to define different control schemes and to evaluate control energies. Table (\ref{table: best_worst}) shows the minimum and the maximum required energy to control studied brain networks and indicates that regional difference in the needed energy is considerable, spanning four orders of magnitude. The regions corresponding to the least and the most required control energy are shown in the table (\ref{table: best_worst}). The frontal-superior gyrus required the minimum cost for controllability, whereas Heschl's gyrus required the maximum control energy. Furthermore, regions that need the minimum control energy have the highest degree in networks, while those regions that require the maximum control energy have the lowest degree. The column $K$ in the table (\ref{table: best_worst}) shows their degree.

The result shows that the required energies for controlling brain networks are various and depend on the input region. So, we then investigate how the network structure and the location of the input region are related to this difference. Specifically, we examine the relationship between the required control energy of input regions and their LCCs in the network. For this purpose, we calculated the LCC of each region in the table (\ref{table: best_worst}), shown in column LCC. According to the outcome, regions that control the networks with the minimum energy have LCC = 3, while regions with the highest control cost have LCC = 5 (except two of them, which have LCC = 4). We determined the smallest and largest achievable LCC in brain networks to evaluate this outcome. The columns $\textrm{LCC}_\textrm{min}$ and $\textrm{LCC}_\textrm{max}$ in the table (\ref{table: best_worst}) show the result demonstrating the minimum and the maximum possible LCC in studied brain networks is three and five, respectively (in two cases the maximum LCC is four). It means that all regions which control the networks with the minimum cost have the smallest LCC, and in contrast, regions that control the brain networks with the maximum energy have the largest possible LCC. The degree of these regions confirms this result since if we consider only one control region, the node with the highest degree has better accessibility (with fewer hops) to other network nodes than the region connected to a small proportion of nodes.

\begin{table}[h]
\centering
\small
\caption{The effect of LCC on the control energy}
\begin{tabular}{c|clcc c | clccc |cc}
\multicolumn{1}{c|}{\multirow{3}{*}{Network}} & \multicolumn{5}{c|}{the minimum control energy} & \multicolumn{5}{c|}{the maximum control energy} & \multicolumn{2}{c}{\multirow{3}{*}{}} \\
& \multicolumn{2}{l}{$\T{Input node}$} & \multicolumn{1}{c}{\multirow{2}{*}{${\varepsilon(t_\T f)}$}} & \multicolumn{1}{c}{\multirow{2}{*}{${K}$}} & \multicolumn{1}{c|}{\multirow{2}{*}{$\T{LCC}$}} & \multicolumn{2}{l}{$\T{Input node}$} & \multicolumn{1}{c}{\multirow{2}{*}{${\varepsilon(t_\T f)}$}} & \multicolumn{1}{c}{\multirow{2}{*}{${K}$}} & \multicolumn{1}{c|}{\multirow{2}{*}{$\T{LCC}$}} & \multicolumn{1}{c}{\multirow{2}{*}{$\T{LCC}_{\T {min}}$}} & \multicolumn{1}{c}{\multirow{2}{*}{$\T{LCC}_{\T {max}}$}}\\
& \multicolumn{1}{l}{Node} & \multicolumn{1}{l}{Region} & & & & Node & Region & & & & & \\ \hline \hline
1 & 2 &Frontal Sup L & 7.4e+11 & 35 & 3 &78& Heschl L & 1.8e+08 & 1 & 5& 3 & 5 \\
2 & 3 &Frontal Sup R & 6.4e+09 & 25 & 3 &78& Heschl L & 1.7e+07 & 1 & 5& 3 & 5 \\
3 & 2 &Frontal Sup L & 3.5e+10 & 36 & 3 &78& Heschl L & 4.3e+07 & 1 & 4 & 3 & 4 \\
4 & 3 &Frontal Sup R & 1.4e+10 & 33 & 3 &62& SupraMarginal L & 7.5e+06 & 2 & 5& 3 & 5 \\
5 & 2 &Frontal Sup L & 4.6e+10 & 45 & 3 &16& Rolandic Oper L & 1.5e+07 & 1 & 4& 3 & 4 \\
6 & 72 &Putamen L & 1.7e+09 & 26 & 3 &16& Rolandic Oper L & 1.6e+06 & 2 & 5& 3 & 5 \\
7 & 73 &Putamen R & 4.8e+11 & 37 & 3 &44& Cuneus L & 5.5e+08 & 3 & 5 & 3 & 5 \\
8 & 66 &Precuneus L & 2.3e+10 & 43 & 3 &10& Frontal Inf Oper L & 5.3e+06 & 1 & 5& 3 & 5 \\
9 & 37 &Hippocampus R & 4.3e+09 & 30 & 3 &20& Olfactory L & 9.1e+06 & 3 & 5& 3 & 5 \\
\hline

\end{tabular}
\label{table: best_worst}
\end{table}

\begin{figure}[!h]
\begin{subfigure}[h]{.49\textwidth}
\centering
\includegraphics[width=.95\linewidth]{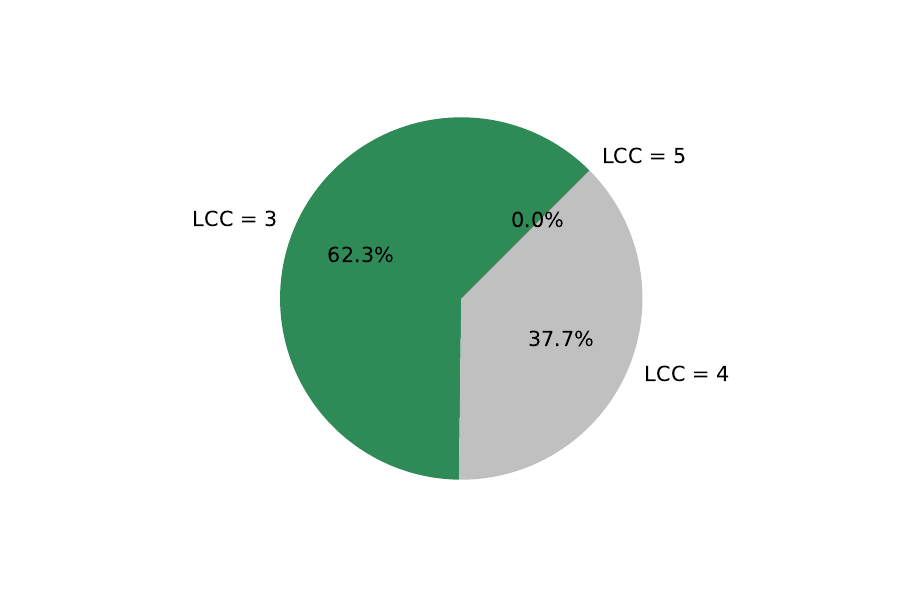}
\caption{regions with the same $\log (\varepsilon(t_\T{f}))$ as the minimum control energy}
\label{fig: The_best_energy}
\end{subfigure}
\begin{subfigure}[h]{.49\textwidth}
\centering
\includegraphics[width=.95\linewidth]{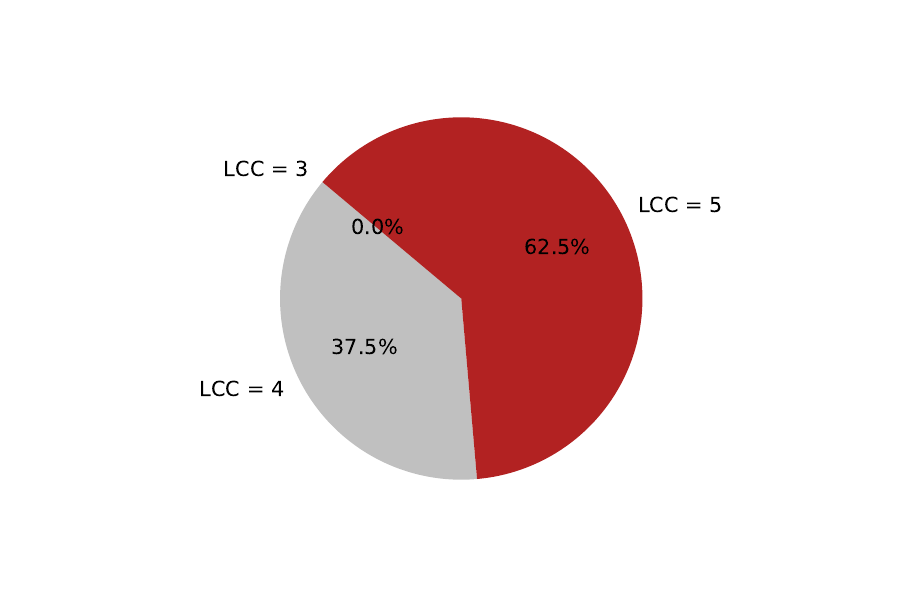}
\caption{regions with the same $\log (\varepsilon(t_\T{f}))$ as the maximum control energy}
\label{fig: The_worst_energy}
\end{subfigure}
\caption{\textbf{Corresponding LCC of regions whose order of energy needed to control brain networks is the same as the minimum or the maximum control energy in studied brain networks. } Regions with the same $\log (\varepsilon(t_\T{f}))$ as the minimum required control energy control the networks mostly with the smallest possible LCC (LCC = 3), and none have LCC = 5. By contrast, regions with the same $\log (\varepsilon(t_\T{f}))$ as the maximum control energy mostly have the largest possible LCC = 5, and none of them control the networks with LCC = 3. 
}
\label{fig: best_worst}
\end{figure}

Note that there are different brain regions whose order of energy needed to control brain networks has the same order with the best and worth control cost. So, in the next step, we determine the LCC of these regions. Figure (\ref{fig: best_worst}) shows the result and confirms the previously obtained outcome. As is shown in this figure, most regions whose order of energy to control brain networks is the same as the order of the minimum control cost have LCC = 3, and none have LCC = 5. The average degree of these regions is 21.2 with the standard deviation std = 6. In contrast, most regions whose order of energy to control brain networks is the same as the order of the maximum control cost have the largest possible LCC = 5, and none have LCC = 3. The average degree of these regions is 1.8, with standard deviation of 0.8.

\begin{figure}[!h]
\begin{subfigure}[h]{.49\textwidth}
\centering
\includegraphics[width=.95\linewidth]{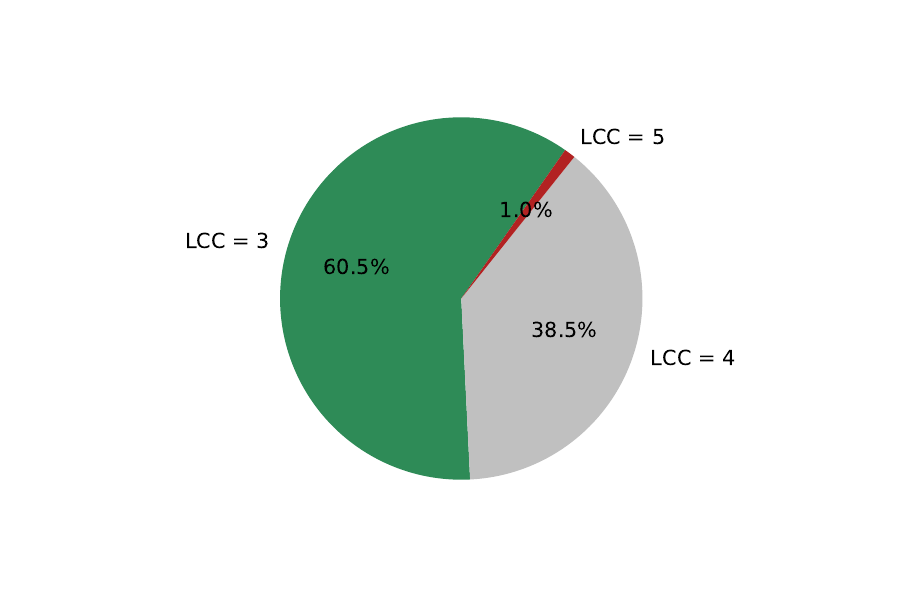}
\caption{low-energy regions}
\label{fig: low-energy regions}
\end{subfigure}
\begin{subfigure}[h]{.49\textwidth}
\centering
\includegraphics[width=.95\linewidth]{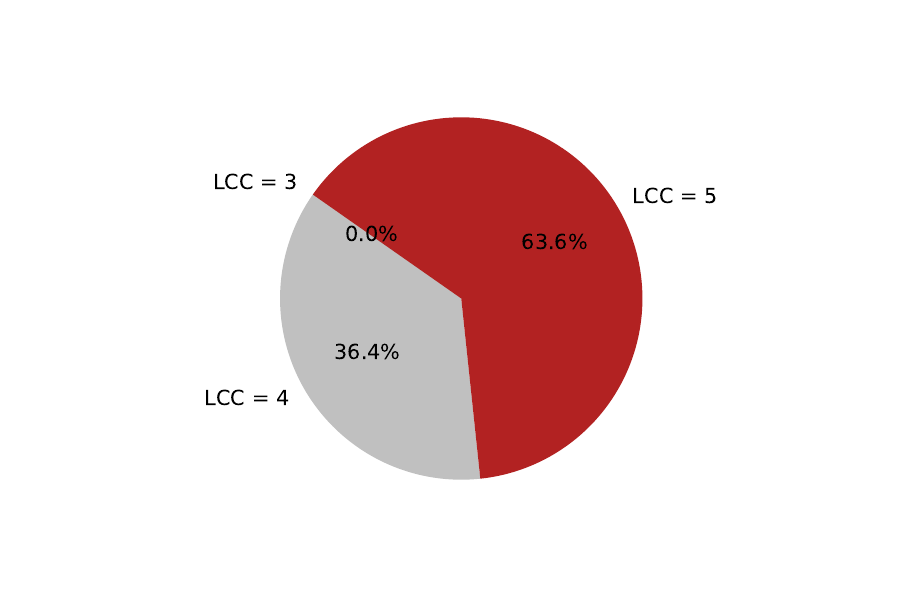}
\caption{high-energy regions}
\label{fig: high-energy regions}
\end{subfigure}
\caption{\textbf{Corresponding LCC of regions that facilitate or constrain brain controllability. } \textbf{(a)} the LCC of regions that facilitate controllability, i.e., regions whose order of control energy is $9 \leq \log ((t_\T{f})) \leq 11$. \textbf{(b)} the LCC of regions that constrain controllability, i.e., regions whose $6 \leq \log (\varepsilon(t_\T{f})) \leq 8$.}
\label{fig: facilitate_constraint}
\end{figure}

We also investigated the role of LCC on a larger scale of regions. Table (\ref{table: best_worst}) shows that $\log (\varepsilon(t_\T{f}))$ in brain networks could be from 6 as the maximum cost to 11 as the minimum order of energy. So, we define two classes of regions according to this range. Region $i$ is low-energy if $i$ controls the network with $9 \leq \log (\varepsilon(t_\T{f})) \leq 11$ otherwise, a high-energy region if its order of control energy is $6 \leq \log (\varepsilon(t_\T{f})) \leq 8$. Hence, low-energy corresponds to regions that facilitate brain controllability, while high-energy regions constrain it. Then, we calculated the LCC of regions corresponding to these classes. Figure (\ref{fig: facilitate_constraint}) shows the result consisting of the earlier outcomes. Figure (\ref{fig: low-energy regions}) shows that most low-energy regions control the network with the minimum achievable LCC. High frequency of these regions (in equal or more than eight brains out of nine) includes Precentral, Frontal Sup, Precuneus, Paracentral Lobule, Frontal Sup Orb, Supp Motor Area, Frontal Sup Medial, Insula, Hippocampus, Putamen, and Parietal Sup. In contrast, most high-energy regions have the largest LCC, and none have LCC = 3, as is shown in figure (\ref{fig: high-energy regions}). The strongest contributors include Frontal Mid Orb, Rolandic Oper, Olfactory, Amygdala, SupraMarginal, and Heschl, and show that Amygdala and Heschl are two regions either appear as isolated nodes or need high energy to control brain networks.

Therefore, the parameter LCC can describe the structural reason for regional differences in the needed energy to control brain networks. The results suggest that the distance between the input region and other regions in brain networks strongly impacts the control cost to steer the brain toward different target states. Regions whose accessibility to other regions is better, i.e., there are fewer hops between them, can control the network with lower energy and are frequently high-degree nodes, while those that have a longer path to reach other regions require more energy to move the brain state between different states and are often low-degree nodes in the network.

\subsection{Consistency with previous findings}

We compare our findings with previous studies that classify brain regions. In these studies, regions with high average controllability are nodes that need low control energy, on average, to steer the network to different target states. The used measure to determine averaged controllability regions is $\text{Trace}({\M W_\textrm{B}(t_\T f)})$. Since we used the same measure, we can compare them with low-energy regions in our study. The result shows that determined low-energy regions include all regions introduced as high in average controllability, plus a few more regions. Therefore, the LCC of these regions in brain networks can be the structural parameter to describe their classification and why these regions need less energy to move the brains around the state space in all directions.

\begin{figure}[!h]
\begin{subfigure}[h]{.49\textwidth}
\centering
\includegraphics[width=0.95\linewidth]{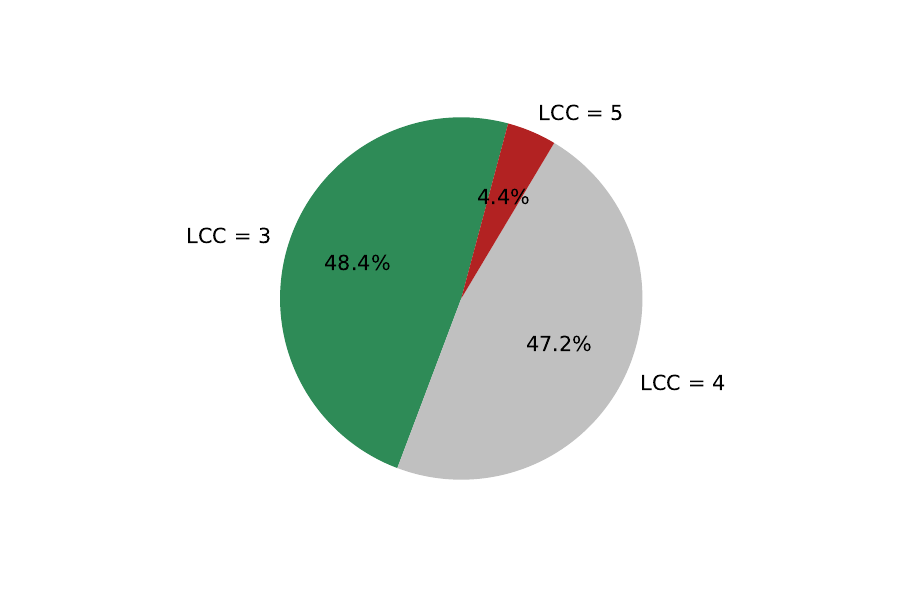}
\caption{Regions associated with the Default mode}
\label{fig: DMN}
\end{subfigure}
\begin{subfigure}[h]{.49\textwidth}
\centering
\includegraphics[width=0.95\linewidth]{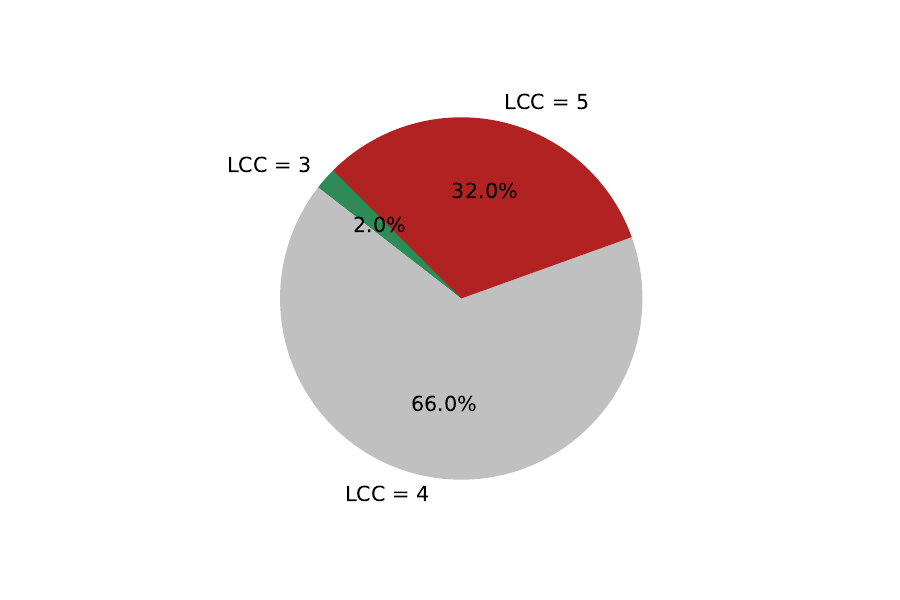}
\caption{Regions associated with the Auditory}
\label{fig: Auditory}
\end{subfigure}
\caption{\textbf{Corresponding LCC of regions that can be mapped to the Default mode or the Auditory networks. } \textbf{(a)} the LCC of regions associated with the DMN. As is shown, most of these regions have a small LCC in controlling brain networks. \textbf{(b)} the LCC of associated regions with the Auditory network. Here, the proportion of regions whose LCCs in controllability of networks are LCC = 5 is more than \textbf{(a)}, and the ratio of regions with LCC = 3 is dramatically less.}
\label{fig: cognitive_systems}
\end{figure}

Additionally, it is shown that regions associated with the default mode network (DMN) contributed the most among averaged controllability regions\cite{gu2015controllability}. By contrast, associated regions with the Auditory and Cingulo-opercular network have the minimum contribution\cite{gu2015controllability}. We hypothesize that a large proportion of areas associated with the DMN should have a small LCC, and by contrast, a small section of associated regions with the Auditory network have LCC = 3. To test these hypotheses, we assigned the corresponding regions of the AAL atlas to the DMN, which includes Frontal Sup Medial, Frontal Med Orb, Cingulum Post, Hippocampus, ParaHippocampal, Angular, and Precuneus, and to the Auditory network that are Rolandic Oper, SupraMarginal, and Heschl (see table (\ref{table: cognitive_systems_regions})). These networks, and the association of regions to them, have previously been extracted from fMRI resting state data \cite{cao2020progressive, long2019psychological}. Figure (\ref{fig: cognitive_systems}) indicates the result. Consistent with our expectation, regions with LCC = 3 comprise a large section of DMN, while regions with LCC = 5 make the smallest part. In contrast, figure (\ref{fig: Auditory}) shows that the proportion of regions with LCC = 3 in the Auditory network is significantly less than DMN.

\begin{table}[h]
\centering
\small
\caption{Regions in the AAL atlas that would be mapped to the Default Mode and Auditory networks}
\begin{tabular}{ll l ll }
\multicolumn{2}{c}{Default Mode} & & \multicolumn{2}{c}{Auditory} \\ \hline
\multicolumn{1}{c}{Node} & \multicolumn{1}{l}{ Region} & & \multicolumn{1}{c}{Node}& \multicolumn{1}{l}{Region} \\ \hline
22, 23 & Frontal Sup Medial & & 16, 17 & Rolandic Oper \\
23, 24 & Frontal Med Orb & & 62, 63 & SupraMarginal \\
34, 35 & Cingulum Post & & 78, 79 & Heschl \\
36, 37 & Hippocampus & & & \\
38, 39	& ParaHippocampal & & & \\
64, 65	& Angular & && \\
66, 67	& Precuneus & && \\
\end{tabular}
\label{table: cognitive_systems_regions}
\end{table}

Therefore, this consistency implies that LCC can describe both the required energy for brain controllability and the needed energy for brain functions. Regions responsible for difficult tasks have larger LCCs in brain networks, while those from which the brain smoothly moves to different states have small LCCs.

\subsection{The impact of the number of paths between regions}

We showed in Sec. (\ref{LCC}) that LCC significantly affects the required energy in controlling brain networks, and we also indicated how this structural parameter determines whether a region facilitates or constrains brain controllability.

Table (\ref{table: best_worst}) shows that the required control energy in brain networks is different with the same LCC and number of input regions. For example, the difference in the minimum control energy spans two orders of magnitude. This section studies what other structural features beyond the LCC affect the control energy. For example, it is shown that increasing the number of paths between inputs and non-input nodes reduces the control energy \cite{klickstein2018control}. However, this factor might be overlooked in our LCC-based analysis when the number of paths between the input and non-input nodes differs in two control schemes with the same LCC.

Therefore, we investigate the controllability of brain networks concerning the needed control energy and the number of paths between the input region and other regions in the brain. However, finding all possible simple paths between any two given nodes in a graph is an NP-hard problem\cite{chatterjee2014novel}, and its order of complexity is $O(N!)$. The worst-case scenario is the complete graph with $N$ nodes and $N!$ simple paths (suppose at the source you have $N-1$ nodes to visit, later $N-2$, $N-3$, ... until you reach the target). Therefore, it is hard to calculate the number of paths from input regions to other regions in brain networks. In this respect, we assess the number of paths between regions by investigating the density of three-node subgraphs named motifs. Motifs are recurrent and statistically significant subgraphs or patterns of a larger graph \cite{milo2002network}. Figure (\ref{fig: Motifs}) in the Appendix shows all possible 3-node motifs in a network. In collected brain networks, we observe only motifs id78 and id238. Increasing the number of these patterns in the network causes the input region to have access to other regions by more separate paths, leading to less needed control energy. For example, any two regions $a$ and $c$ in 3-node motif id238 can reach each other by the direct link $(a, c)$ or by the third region $b$ and the path $(a, b), (b, c)$. Therefore, in the following, we investigate control energy in brain networks concerning the number of motifs in their structures.

For this purpose, we first determine the number of motifs id78 and id238 in studied brain networks that are shown in table (\ref{table: motifs}) by $N_\T{m}$. Networks in this table are sorted in descending order based on the number of motifs in their structure so that network 1 has the most and network 6 has the minimum motifs. Then, we compare the control energy of regions with the same LCC in these networks. We first considered LCC = 3 and calculated the average control energy of all regions that control the networks with LCC = 3. The result is shown in column $\bar{\varepsilon}_\T{LCC = 3}(t_\T{f})$ indicating regions with the same LCC (LCC = 3) can control brain networks with less control energy if they have more motifs in their structures. It means that even with the smallest possible LCC, brain networks need different energies that could make controllability impossible. Likewise, we compared the control energy of regions with LCC = 5 in column $\bar{\varepsilon}_\T{LCC = 5}(t_\T{f})$. Note that the maximum possible LCC in two networks is four, so we did not consider networks 3 and 5 in this part of the study. The result shows the same outcome, and regions with LCC =5 need more energy to control brain networks with fewer motifs. Therefore, the pattern of white matter fibers between regions in brain networks, evaluated by motifs, affects the number of paths between regions, and the higher number of these patterns, the less energy is needed to control.
\begin{table}[h]
\centering
\small
\caption{The relation between the number of motifs in brain networks and the needed control energy}
\begin{tabular}{c| c | c c | c}
\multicolumn{1}{c|}{\multirow{1}{*}{Network}} & \multicolumn{1}{c|}{\multirow{1}{*}{$N_\text{m}$}} & \multicolumn{1}{c}{\multirow{1}{*}{$\bar{\varepsilon}_\T{LCC = 3}(t_\T{f})$}} & \multicolumn{1}{c|}{\multirow{1}{*}{$\bar{\varepsilon}_\T{LCC = 5}(t_\T{f})$}} & \multicolumn{1}{c}{\multirow{1}{*}{$\bar{\varepsilon}(t_\T{f})$}}
\\ \hline \hline
1 & 7433 & 2.5e+11 & 5.0e+09 & 1.6e+11 \\
7 & 6447 & 1.3e+11 & 1.3e+09 & 7.1e+10 \\
8 & 6329 & 5.1e+09 & 1.3e+07 & 3.2e+09 \\
2 & 6006 & 3.3e+09 & 5.2e+08 & 2.2e+09 \\
4 & 5920 & 3.3e+09 & 3.1e+08 & 2.2e+09 \\
9 & 5607 & 2.3e+09 & 1.1e+08 & 1.1e+09 \\
6 & 5262 & 9.7e+08 & 4.9e+07 & 4.3e+08 \\
\hline
\end{tabular}
\label{table: motifs}
\end{table}

\begin{figure}[h]
\begin{subfigure}[h]{.4\textwidth}
\centering
\includegraphics[width=1\linewidth]{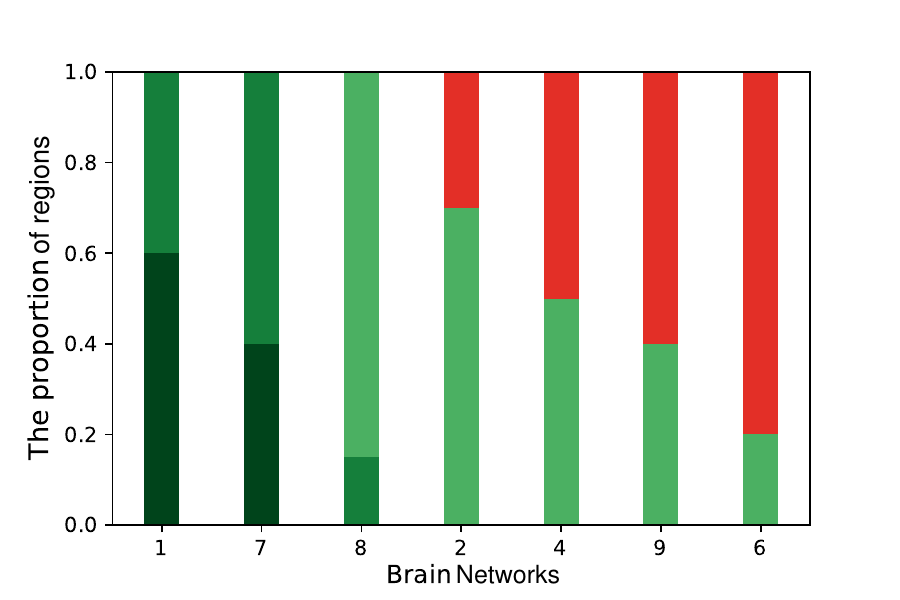}
\caption{The control energy of regions with LCC = 3}
\label{fig: stacked_bar_chart(a)}
\end{subfigure}
\begin{subfigure}[h]{.4\textwidth}
\centering
\includegraphics[width=1\linewidth]{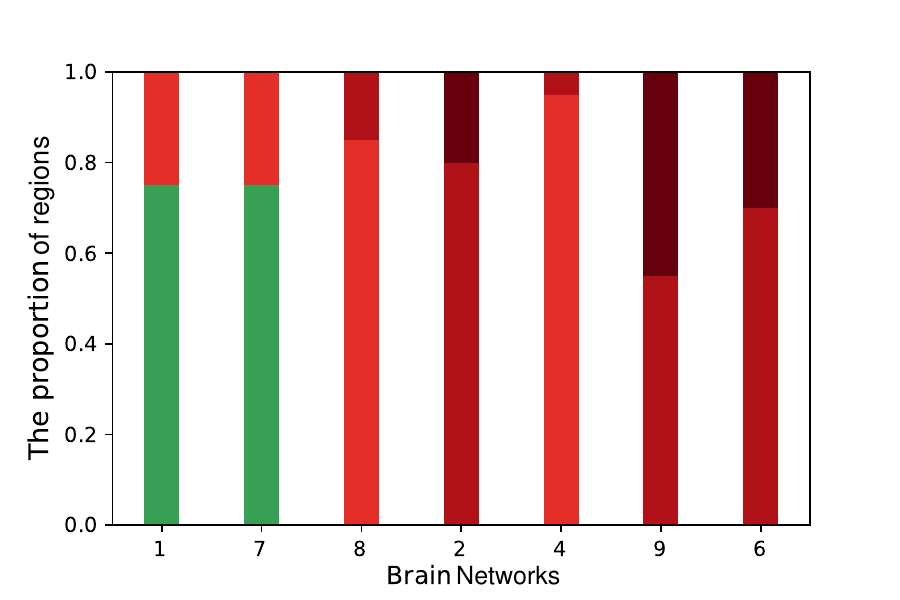}
\caption{The control energy of regions with LCC = 5}
\label{fig: stacked_bar_chart(b)}
\end{subfigure}
\begin{subfigure}[h]{.19\textwidth}
\centering
\includegraphics[width=1.4\linewidth]{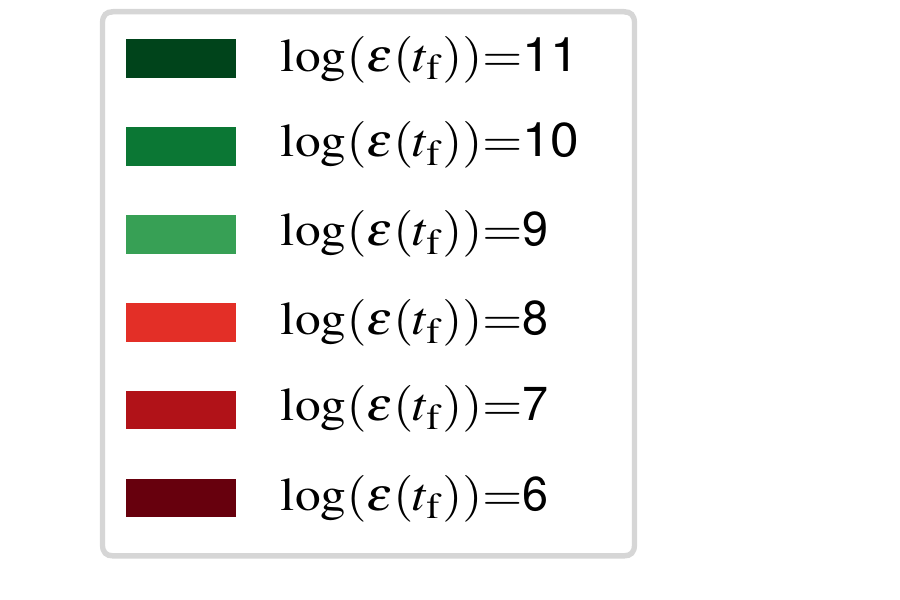}
\end{subfigure}
\caption{\textbf{The control energy of regions with the same LCC in different brain networks. } Networks are sorted based on the number of motifs in their structure in descending order. \textbf{(a)} The control energy of regions that control networks with LCC = 3. Decreasing the number of motifs leads to needing more control energy even with the same LCC. \textbf{(b)} The control energy of regions that control networks with LCC = 5. We have the same outcome in such a way that in network 1, regions with LCC = 5 control the network with $\log (\varepsilon(t_\T{f})) = 9$ or 8, while regions with the same LCC in network 6 control the network with $\log (\varepsilon(t_\T{f})) = 6$ or 7. }
\label{fig: stacked_bar_chart}
\end{figure}

We also provided the detail of the required energy of regions with the same LCC in each network. Figure (\ref{fig: stacked_bar_chart(a)}) shows the control energy of regions with the LCC = 3. As is shown, these regions in network 1, which have the maximum number of motifs, control the network with mostly $\log (\varepsilon(t_\T{f})) = 11$. While in network 6, which has the minimum number of motifs, regions with the same LCC = 3 control the network with $\log (\varepsilon(t_\T{f})) = 8$. The difference between them is three orders of magnitude. Figure (\ref{fig: stacked_bar_chart(b)}) shows the same comparison but for regions with LCC = 5. Here, there is the same result. Note that the number of links in the studied networks is almost the same, and the little difference between them could not significantly impact the number of motifs. For example, although network 4 has fewer links than network 6, but has more motifs in its structure.

\section*{Discussion}
Although the brain networks are shown to be theoretically controllable by one control region, in practice, these networks are extremely hard to control \cite{gu2015controllability,karrer2020practical}. One may consider additional input nodes to reduce the control energy, but the exact number of required nodes and their locations could still be unknown to decrease control energy efficiently. We employed the suggested approach in our prior work to find how many extra input nodes, at minimum, are required to reduce LCC in brain networks. For this purpose, we considered the minimum LCC = 1. Finding these regions could provide more insight into regions that control brain networks with smaller energy. Table (\ref{table: LCC=1}) shows the result, and parameter $N_\T{i}^{1}$ indicates the required input nodes to control the network with LCC = 1. Note that the solution for this constraint on LCC is not unique. So, we determined one hundred different input sets to find which regions have the most participation to address this constraint. Regions that participated in all input sets have a primary role in reducing the LCC and energy in controlling brain networks. These regions are shown in column $\T{R}_\T{LCC=1}$. Furthermore, analyzing repeated regions indicates that Precuneus and Temporal Mid are repeated in $66\%$ and $45\%$ of brain networks, respectively, as the most contributors.
\begin{table}[h]
\centering
\small
\caption{Controlling brain networks with LCC = 1 }
\begin{tabular}{c|lll}
\multicolumn{1}{c|}{Networks} & \multicolumn{1}{c}{$N_\textrm{d}^{1}$} & $\T{R}_\T{LCC=1}$       \\
\hline  \hline
1          & 9   &  Frontal Inf Tri L,  Frontal Med Orb L, Precuneus L, Temporal Pole Sup L, Temporal Pole Sup R               \\  
2          & 10   & Calcarine R, Precuneus L, Temporal Mid L, Temporal Pole Mid R                    \\
3          & 8     & Hippocampus R, Occipital Mid R,  Postcentral R         \\    
4          & 8     & Frontal Sup R, Frontal Sup Medial L, Lingual L, Precuneus L, Temporal Mid L         \\    
5          & 10   & Frontal Sup Medial L, Precuneus L, Thalamus L, Temporal Mid R           \\
6          & 11   & Hippocampus R, Occipital Mid L, Postcentral R, Parietal Inf L, Precuneus L                   \\
7          & 10   & Frontal Sup Orb L, Temporal Inf R                     \\
8          & 10   & Precuneus L                   \\
9          & 8     & Insula L, Cingulum Ant L,  Hippocampus R, Fusiform L                 \\
                              \hline
\end{tabular}
\label{table:  LCC=1}
\end{table}

\section*{Conclusion}
The controllability of brain networks is the ability to steer the brain from its initial state to any desired one by changing the neurophysiological activity of a region as an input region. Controlling the dynamics of brain networks has many uses, like medical purposes and therapeutic processes. The energy needed to control a network is a central metric to quantify the difficulty of the controllability, and the control energy of brain networks is considerably different depending on the selected input region. Employing our prior work about the input node placement and its effect on the longest control chain (LCC) in the controllability of networks, we indicate the role of the structure of white matter fibers and the location of input regions in the required energy of controlling brain networks. Specifically, we show a strong relation between the cost of brain network controllability and the LCC of the input region, i,e, the longest distance between the input region and other regions in the network. Therefore, regions with smaller LCCs can move the brain toward different target states with lower energy, while those with larger LCCs need higher control energy to steer the brain around the state space. Furthermore, we show that the connection of white matter fibers between regions makes specific motifs in the architecture of brain networks that increase the number of paths between the input region and other brain regions. We show that the higher number of these patterns in brain networks causes less energy to be needed for controllability.

\subsection*{Author contributions}

S.A., A.G. conceived the project. S.A., A.G., A.F. designed the research. S.A. performed simulations and analyzed the
empirical data. S.A., A.F., A.G. wrote the manuscript.

\subsection*{Acknowledgments}
This research has been conducted using the Iranian Brain Mapping Biobank (IBMB) Resource. We would like to thank Dr. Batouli for sharing the dataset. A. Ghasemi gratefully acknowledges funding from the Alexander von Humboldt Foundation for his research fellowship at the University of Passau, Germany.

\subsection*{Data availability}
The IBMB data are available from http://ibmb.nbml.ir/ on reasonable request. The data used during this study are also available from the corresponding author, upon reasonable request.

\section*{Appendix} \label{appendix}

\begin{figure}[!h]
  \centering
  \includegraphics[width=1\linewidth]{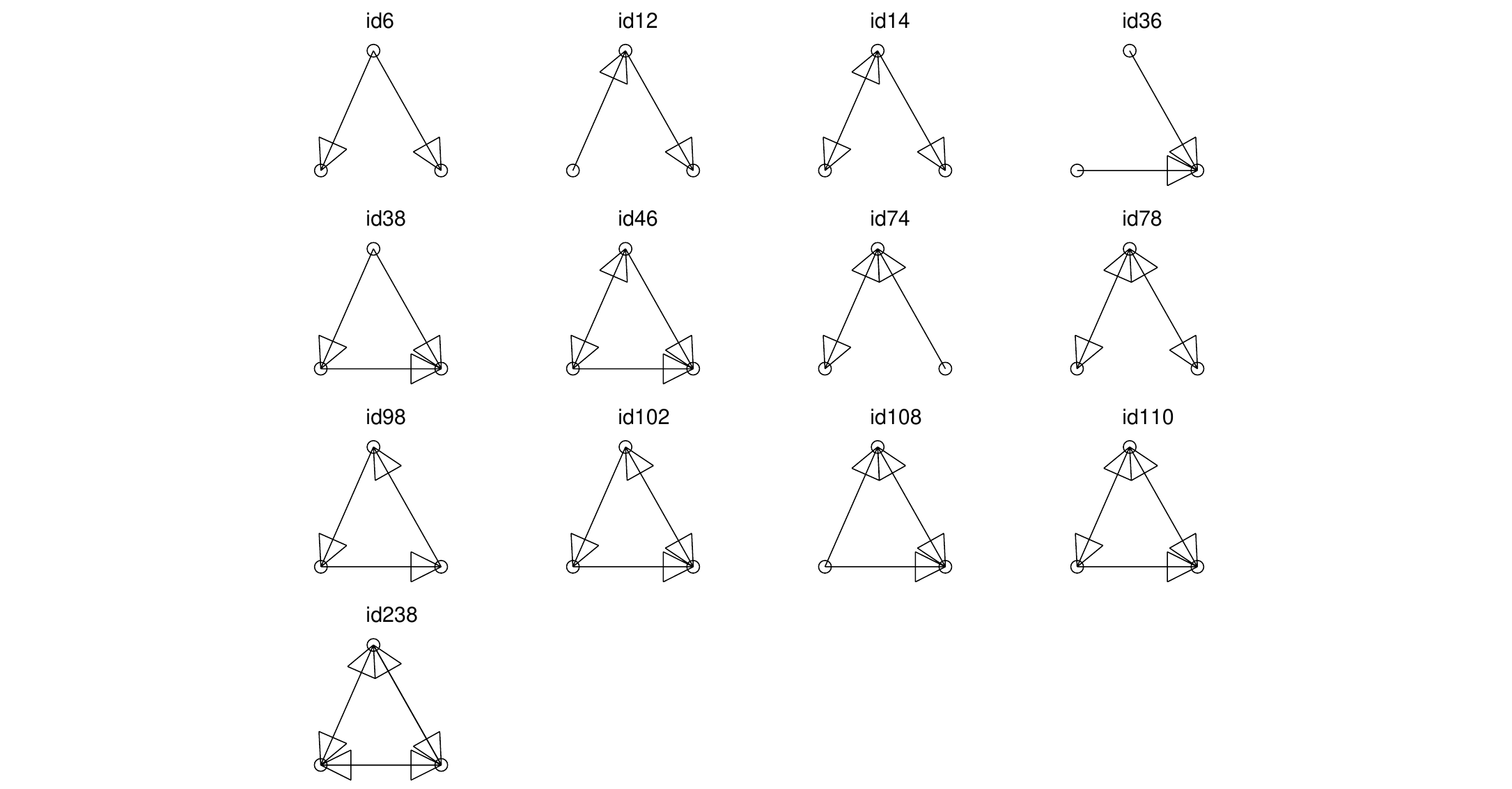}  
\caption{\textbf{All possible subgraphs with three nodes. } The only motifs in brain networks are id78 and id238, according to obtained data from brain communications and two-way links between regions.}  
\label{fig: Motifs}
\end{figure}

\bibliography{refs}

\end{document}